\documentclass[pdflatex,sn-mathphys-num]{sn-jnl}% Math and Physical Sciences Numbered Reference Style 
\usepackage{graphicx}%
\usepackage{multirow}%
\usepackage{amsmath,amssymb,amsfonts}%
\usepackage{amsthm}%
\usepackage{mathrsfs}%
\usepackage[title]{appendix}%
\usepackage{xcolor}%
\usepackage{textcomp}%
\usepackage{manyfoot}%
\usepackage{booktabs}%
\usepackage{algorithm}%
\usepackage{algorithmicx}%
\usepackage{algpseudocode}%
\usepackage{listings}%
\usepackage{float} 
\usepackage{placeins}
\theoremstyle{thmstyleone}%
\theoremstyle{thmstyletwo}%

\theoremstyle{thmstylethree}%

\begin{document}

\title[Deep Learning for Automated Wound Classification And Segmentation]{Deep Learning for Automated Wound Classification And Segmentation}


\author*[1]{\fnm{Md. Zihad Bin} \sur{Jahangir}}\email{zihad.bscincse@gmail.com}

\author[2]{\fnm{Sumaiya} \sur{Akter}}\email{sumaiya.akter.cse@ulab.edu.bd}

\author[3]{\fnm{MD Abdullah Al} \sur{Nasim}}\email{nasim.abdullah@ieee.org}

\author[4]{\fnm{Kishor Datta} \sur{Gupta}}\email{kgupta@cau.edu}

\author[5]{\fnm{Roy} \sur{George}}\email{george@cau.edu}


\affil*[1]{\orgdiv{Department of Computer Science and Engineering}, \orgname{Southeast University}, \city{Dhaka 1215}, \country{Bangladesh}}

\affil[2]{\orgdiv{Department of Computer Science and Engineering}, \orgname{University of Liberal Arts Bangladesh}, \city{Dhaka 1207}, \country{Bangladesh}}

\affil[3]{\orgdiv{Research and Development Department}, \orgname{Pioneer Alpha}, \city{Dhaka 1205}, \country{Bangladesh}}

\affil[4, 5]{\orgdiv{Department of Computer and Information Science}, \orgname{Clark Atlanta University}, \city{Atlanta, GA 30314}, \country{USA}}


\abstract{Wounds, such as foot ulcers, pressure ulcers, leg ulcers, and infected wounds, come up with substantial problems for healthcare professionals. Prompt and accurate segmentation is crucial for effective treatment. However, contemporary methods need an exhaustive model that is qualified for both classification and segmentation, especially lightweight ones. In this work, we tackle this issue by presenting a new architecture that incorporates U-Net, which is optimized for both wound classification and effective segmentation. We curated four extensive and diverse collections of wound images, utilizing the publicly available Medetec Dataset, and supplemented with additional data sourced from the Internet. Our model performed exceptionally well, with an F1 score of 0.929, a Dice score of 0.931 in segmentation, and an accuracy of 0.915 in classification, proving its effectiveness in both classification and segmentation work. This accomplishment highlights the potential of our approach to automating wound care management.}

\keywords{Healthcare, Wound Segmentation and Classification, Deep Learning, Deep Convolutional Neural Network}



\maketitle

\section{Introduction}\label{sec1}

The rising number of chronic wounds is now a serious global issue, with important implications for both healthcare and financial systems.  Diabetic ulcer treatments alone are estimated to exceed \$50,000 \cite{han2017chronic}, while the management of chronic wounds is expected to surpass \$25 billion per year \cite{goldstein2007cholinergic}. In addition, persistent wounds sometimes need prolonged healing periods, which might endure for several years.  Throughout the healing process, continual monitoring and assessment of the wound by a doctor are crucial for the success of the treatment and therapeutic progress \cite{nguyen2017spatial}. A variety of methods are used in the process of monitoring and recording the rate of wound healing \cite{li2017deep}. Integral to these methods is the segmentation algorithm which has a paramount role in the evaluation process since it is responsible for outlining and categorizing the wound images \cite{kumar2017dataset}. A precise evaluation relies on proper segmentation. Therefore, a dependable algorithm is critically required to accurately segment and classify wound regions in images, enabling continuous monitoring and assessment.

Given the aforementioned challenges that are associated with the management of chronic wounds, the present study aims to pursue two objectives. We aim to develop a unified model capable of performing segmentation and classification tasks simultaneously. The main goal of incorporating this information is to simplify the treatment plan and offer help with wound evaluation. In addition, we want to improve the model's computational efficiency by minimizing its computing footprint and resource needs.  Optimizing this aspect is particularly crucial in resource-constrained environments.

The significance of the present study lies in the development of a modified U-Net architecture specifically tailored to meet the objectives of the investigation. This modification demonstrates enhanced efficacy in both classification and segmentation tasks, guaranteeing more precise and streamlined evaluations of wounds. We introduce a novel strategy that integrates recent developments in deep learning with well-established techniques to resolve the challenges pertaining to the treatment of chronic wounds. Our goal is to promote enhanced patient results and enhance healthcare effectiveness through novel technology solutions.

\section{Literature Review}\label{sec2}

The integration of various neural network types, notably Convolutional Neural Networks (CNN) \cite{zaheer2018gpu}, You Only Look Once (YOLO) \cite{redmon2016you} for object detection, and U-Net \cite{ronneberger2015u} for segmentation, has significantly advanced the study of wound care. The U-Net architecture, based on the CNN principles and defined by a simplified encoder-decoder design, has become the standard segmentation tool in medical image segmentation and is regarded as state-of-the-art architecture \cite{litjens2017survey} That versatility and adaptability have since given rise to an array of different versions of the U-Net framework such as Connected-UNets \cite{baccouche2021connected} and AU-Net \cite{sun2020aunet} designed to address specific optimization difficulties. These models have shown impressive performance in a variety of applications in medical image processing, including object segmentation, detection, and classification.

Alongside those developments in segmentation, the RCNN family of models has driven the state of the art in object detection and segmentation in computer vision. RCNN - Girshick et al. introduced the RCNN object detection framework, which achieved a big breakthrough in 2014 through a two-step process to generate region proposals and then CNN classification for each region. Although accurate, RCNN was computationally expensive due to evaluating each region proposal independently \cite{girshick2014rich}. Girshick (2015) presented a method, namely Fast RCNN, that aimed to overcome these inefficiencies by utilizing a convolutional neural network (CNN) to process the entire image and generate a feature map. This method significantly decreased computation time and incorporated a multi-task loss that merged bounding box regression and classification, thereby improving performance \cite{girshick2015fast}. Ren et al., in their 2016 paper on Faster R-CNN, introduced a Region Proposal Network (RPN) that shared convolutional layers with the object recognition network. This innovation accelerated both speed and accuracy through an end-to-end system that simultaneously predicted objectness scores and object boundaries \cite{ren2015faster}. He et al. introduced Mask RCNN in 2017, which expanded upon Faster RCNN by adding a separate component for predicting segmentation masks, thereby enabling instance segmentation \cite{he2017mask}.

Recent progress, such as the extensive use of deep convolutional neural networks, has notably focused on addressing the intricacies of medical image processing, particularly in the field of wound segmentation.  A novel method TransMix \cite{kuo2024improving}, combines LCF and AGP approaches to provide a powerful training framework for segmentation models for diabetic foot ulcers. This strategy improves training stability, especially in situations where the data isn't well annotated. FUSegNet \cite{dhar2024fusegnet} presents an automated approach for segmenting diabetic foot ulcers into distinct parts. The system integrates an EfficientNet-b7 backbone, featuring a customized scSE module operating in parallel across spatial and channel dimensions, incorporating numerous enhancements that surpass current methodologies. HarDNet-DFUS \cite{liao2022hardnet} was created to aid in differentiating between medical images of diabetic foot ulcers and colonoscopy polyps, and it performs exceptionally well in both tasks. It achieved the first position in the Diabetic Foot Ulcer Segmentation Challenge (DFUC2022) at MICCAI 2022, obtaining a mean Dice score of 0.7287. Additionally, it displays efficient segmentation of colonoscopy polyps. Also, Reference \cite{ali2022lightweight} showed an early plan for dividing foot ulcers into groups. According to their study, this plan did better than all other segmentation algorithms in the Foot Ulcer Segmentation Challenge 2021 dataset, which is a big step forward in the field. Work \cite{kendrick2022translating} introduces DFUC2022, the largest segmentation dataset of diabetic foot ulcers featuring manual annotations by medical professionals. The authors present an enhanced FCN32 VGG network that surpasses the existing benchmark by attaining a Dice score of 0.7446. This achievement places them at the top of the DFUC2022 challenge scoreboard.

Several studies have been conducted in recent years to implement AI in the medical sector \cite{jahangir2023introduction, rahman2024breast}. Wang et al. (2020) presented a lightweight, effective wound segmentation model, using MobileNetV2 and linked component labeling that achieves performance on par with deeper networks. Having trained on 1109 foot ulcer pictures, their model emphasizes the need for precise wound measuring in both diagnosis and treatment \cite{wang2020fully}. Rostami et al. (2021) presented a deep learning-based method, using a fine-tuned AlexNet model for burn wound picture classification. Their classifier, which divides wounds into many classifications depending on circumstances, showed an over 90\% accuracy increase over earlier approaches, therefore indicating the possibility for improved diagnosis and treatment in wound care \cite{rostami2021multiclass}. Deep convolutional neural networks and support vector machines enable Chauhan and Goyal's (2020) BPBSAM, a body part-specific burn severity assessment model. With an accuracy of up to 84.8\% and an average F1 score of 77.8\% on many datasets, BPBSAM notably improved burn severity classification by using non-burn body component photos for training. This innovative approach addresses the challenges of limited burn image data, significantly enhancing burn diagnosis and potentially aiding in burn region segmentation \cite{chauhan2020bpbsam}.

The use of sophisticated neural network structures such as the RCNN lineage, the U-Net lineage, and innovative techniques for wound segmentation have revitalized the field of medical image analysis. This approach emphasizes ongoing advancements in deep learning for WDS, enhancing the accuracy and efficacy of automated wound contour identification while advancing wound care management automation.

\section{Methodology}\label{sec3}
% Insert the image
\begin{figure}[H]
    \centering
    \includegraphics[width=0.5\textwidth]{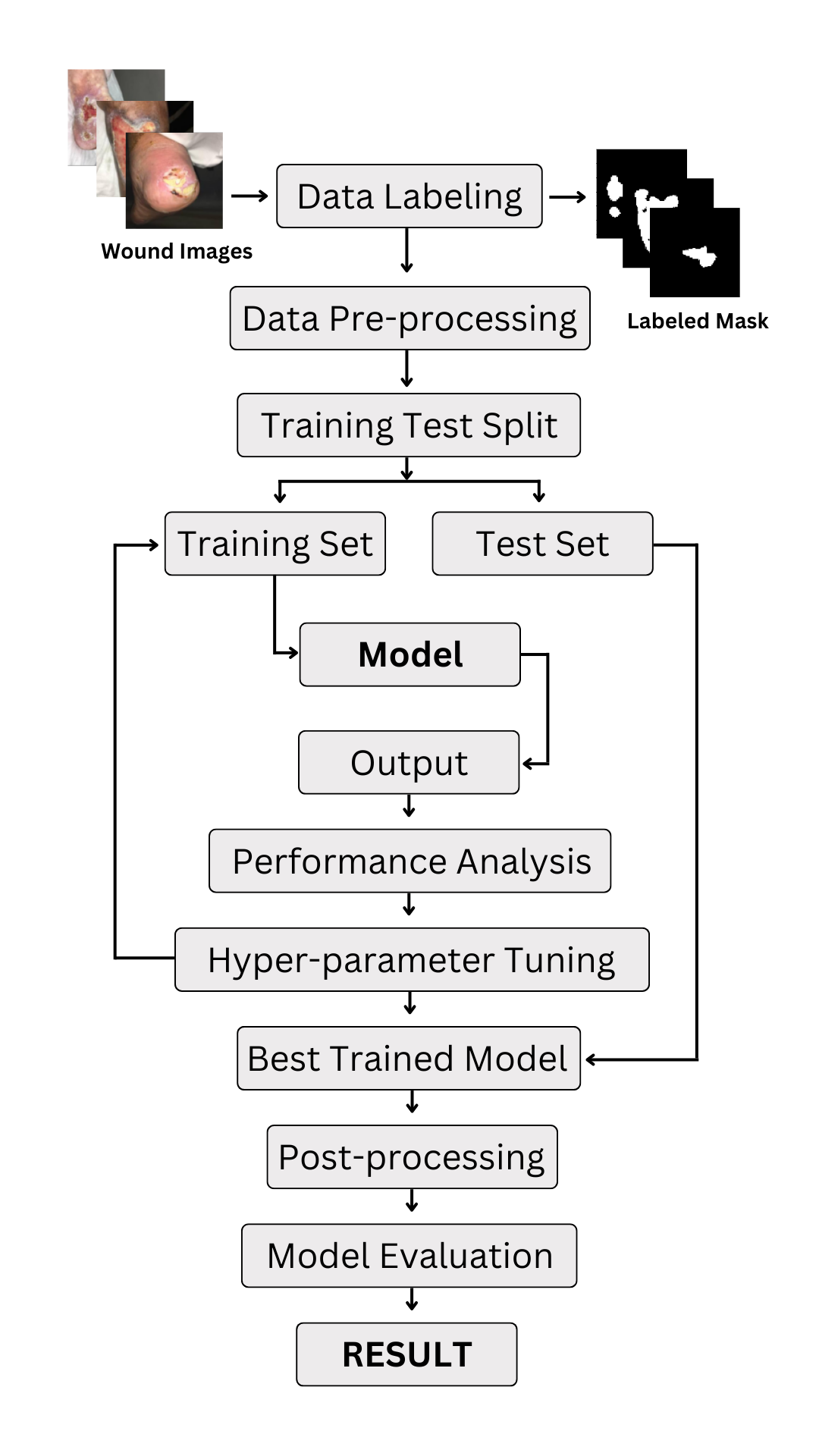}
    \caption{A succinct graphical representation illustrating the study's methodology.}
    \label{fig:methodology_image}
\end{figure}

This section aims to provide an outline of the methodology part, which includes many levels designed to develop the best strategies for image analysis, which involve wound image classification as well as segmentation. Figure \ref{fig:methodology_image} is a diagram illustrating the process of conducting the study, with emphasis on the processes of data preparation, model training, and analysis.

\subsection{Dataset and Preprocessing}
Within our dataset, we categorized foot ulcers, pressure ulcers, leg ulcers, and infected wounds as four distinct categories of wounds. In addition to the publicly accessible Medetech dataset \cite{medetecwounddatabase}, the dataset was further enriched with new photos acquired from internet browsing. The categories are displayed in Table \ref{tab:wound_counts}. The data to be used for training was pre-processed by standardizing their sizes to fit within a common window size of 128 × 128 pixels while maintaining color information. During resizing, all images had to be made homogeneous to make every computation effective while training.

\begin{table}[ht]
\centering
\caption{Dataset breakdown}
\label{tab:wound_counts}
\begin{tabular}{|l|c|}
\hline
Type            & Image count \\
\hline
Infected Wound  & 51          \\
Foot Ulcer      & 52          \\
Leg Ulcer       & 63          \\
Pressure Ulcer  & 90          \\
\hline
\end{tabular}
\end{table}

% Insert the image
\begin{figure}[H]
    \centering
    \includegraphics[width=1\textwidth]{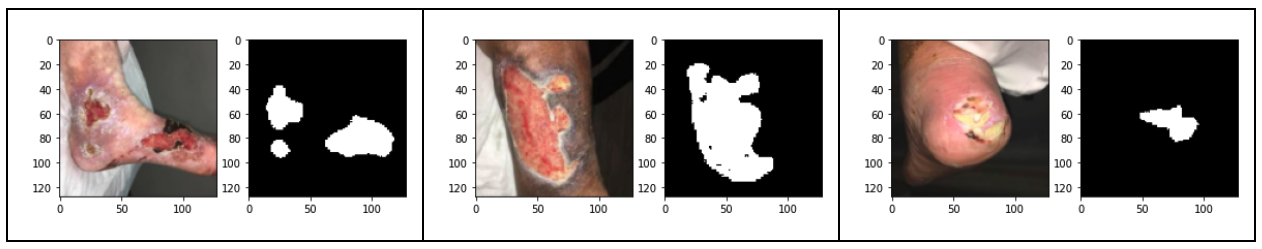}
    \caption{The annotated dataset utilized for model development and assessment is illustrated by examples displaying wound images alongside their segmented masks}
    \label{fig:sample_data}
\end{figure}

This is because we segmented each image manually using the Make Sense AI tool to ensure accurate labeling for training purposes (Source: \url{https://www.makesense.ai/}). The exact definition of wound sites was ensured by this careful segmentation technique resulting in efficient learning from labeled data by the model. Figure \ref{fig:sample_data} shows samples from our training and evaluation set representing wound pictures along with their respective segmented masks.

\subsection{Model Architecture}
\begin{figure}[H]
    \centering
    \includegraphics[width=1\textwidth]{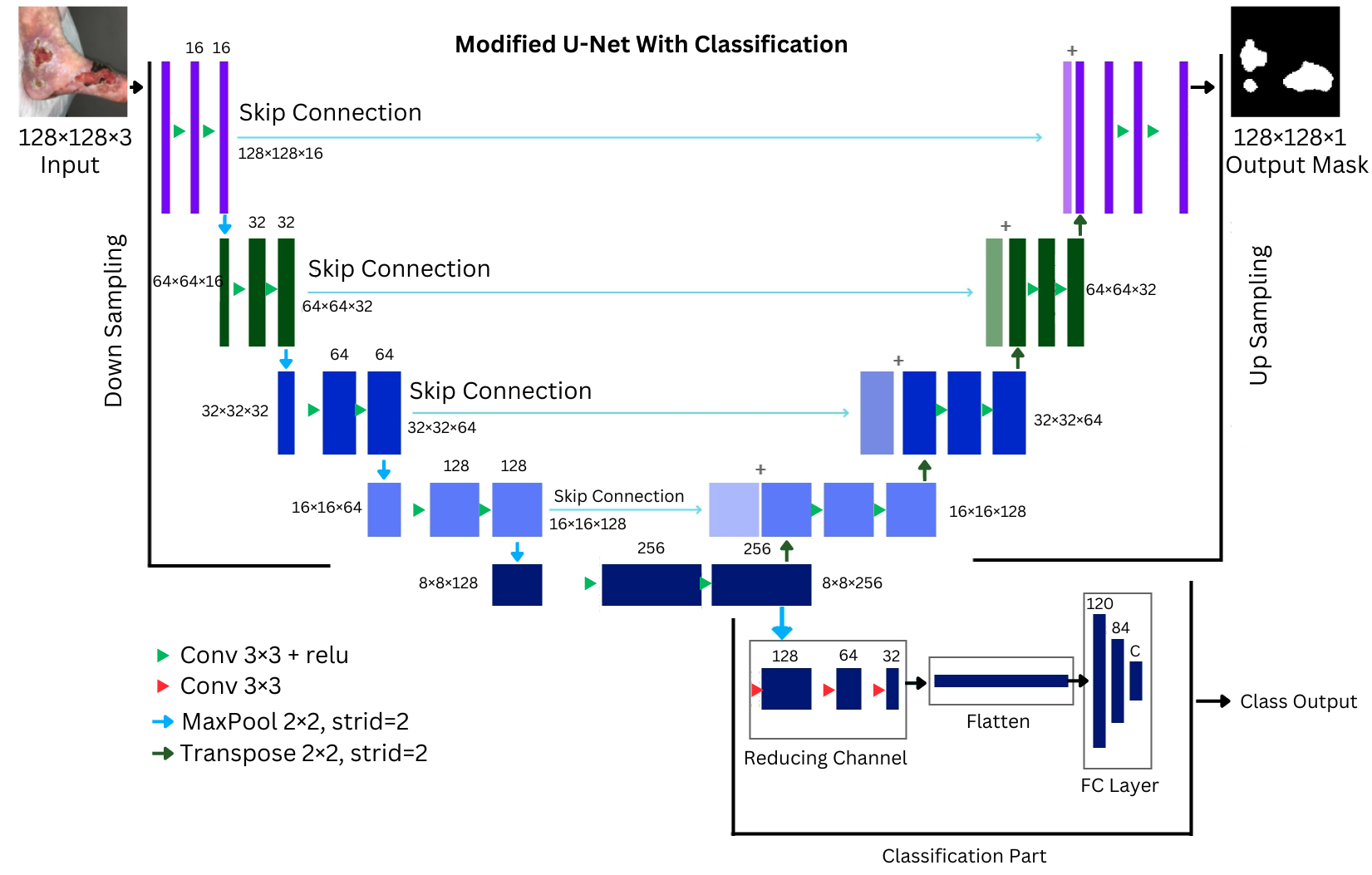}
    \caption{Proposed Architecture (Modified U-Net)}
    \label{fig:model_architecture}
\end{figure}

Figure \ref{fig:model_architecture} illustrates the architecture we suggest which is based on the U-Net architecture and designed to do effective classification and segmentation of wounds. Our model takes images sized 128×128×3 as input. We created a much smaller, faster model by making various changes to the basic structure of U-Net thereby customizing it strictly for our purpose. To decrease computational complexity, our model operates on 128×128 images rather than the primary U-Net's 572×572 images. We incorporate zero padding in each convolutional layer, differing from the original U-Net, which does not use padding. For downsampling, we apply a max pool of 2×2 operation with a stride of 2. Our system's architecture is divided into 3 primary components: downsample, upsample, and classification.

% Insert the image

The downsample part contains 5 blocks in all with each block containing 2 convolution layers, along with a ReLU activation function, followed by a max pool layer.  Each convolution layer used a kernel size of 3×3 with zero padding. The convolutional layers begin with 16 filters, doubling with every block through 16,32,64,128, and finally 256 filters respectively. The max-pool layers are characterized by a stride of 2, a size of 2×2. Hence as it passes through five downsampling blocks, the size of the feature map reduces to be equal to 8×8×256.

Following the downsampling section, we introduce a classification section. This section starts with three consecutive convolutional layers, each with a 3×3 kernel size, aimed at reducing the number of channels. The number of channels decreases from 256 to 128 in the first convolutional layer, 128 to 64 in the second layer, and 64 to 32 in the third convolutional layer, resulting in an 8×8×32 output. This is then followed by a flattening procedure which allows us to make use of three ‘fully connected’ layers. The first fully connected layer has nodes numbering up to 120 while the second number has 84 and the last one possesses 4 nodes for all wound classes respectively.

The output of the downsampling part is concurrently passed to the upsample part to generate segmentation masks. The architecture of this part is similar to the downsampling path, but it uses transposed convolutions for upsampling. For example, the upsampling section has four up-sampling blocks. Where each block reduces the channel number and raises the spatial dimensions, going from size 8×8×256 to 16×16×128, then to size 32×32×64, followed by size 64×64×32, and finally producing a segmentation mask of size 128×128×1. Each upsampled block consists of two convolutional layers with ReLU activation and zero padding followed by a transpose convolutional layer with kernel size equal to (2×2) and stride equal to (2×2).

To maintain the spatial information, the concept of skip connection is implemented between the equivalent downsample and upsample blocks. These skip connections maintain fine-grained details across the network by merging feature maps from the descending path with those from the equivalent ascending path.

\subsection{Training}
We trained our model for 500 epochs on an NVIDIA GTX1050 GPU with 4GB memory. Our architecture had two outputs; hence, we used two different loss functions to optimize the model. For classification output, "Cross-Entropy Loss" was used, and "BCE With Logits Loss" was used in mask prediction. The total loss or the final loss is calculated as the sum of the classification loss and the segmentation loss.

\textbf{Cross-Entropy Loss:} Often used in multi-class classification problems, measures the performance of a classification model whose output is a probability distribution Equation (\ref{eq:cross_entropy_loss}). The loss increases as the predicted probability diverges from the actual label. Where \( p \) is the predicted probability distribution, \( y \) is the true label, \( C \) is the number of classes, \( p_c \) is the predicted probability for class \( c \), and \( y_c \) is the true probability for class \( c \).

\begin{equation}
\label{eq:cross_entropy_loss}
\text{CrossEntropyLoss}(p, y) = -\sum_{c=1}^{C} y_c \log(p_c)
\end{equation}

\textbf{BCE With Logits Loss:} Commonly used in binary classification tasks where the output is a probability distribution Equation (\ref{eq:bce_with_logits_loss}). It combines a sigmoid layer and the binary cross-entropy loss in a single class. Where \( x \) is the raw output, and \( y \) is the true label.

\begin{equation}
\label{eq:bce_with_logits_loss}
\text{BCEWithLogitsLoss}(x, y) = -\left[ y \cdot \log\left( \frac{1}{1 + e^{-x}} \right) + (1 - y) \cdot \log\left( 1 - \frac{1}{1 + e^{-x}} \right) \right]
\end{equation}

\textbf{Total Loss:} Sum of "Cross-Entropy Loss" with "BCE With Logits Loss" Equation (\ref{eq:total_loss})

\begin{equation}
\label{eq:total_loss}
\text{TotalLoss} = \text{CrossEntropyLoss}(p, y) + \text{BCEWithLogitsLoss}(x, y)
\end{equation}

Upon computing the total loss, backpropagation is executed to refine the model weights. During the training phase, we meticulously saved the model state that yielded the lowest total loss, thus ensuring the retention of the optimal model parameters.

\subsection{Post-processing}
Once we had the anticipated mask from the model, we implemented a post-processing procedure to transform the grayscale mask into a binary format. The predicted masks are initially in grayscale, representing probabilities. In order to convert these into binary masks that are appropriate for additional analysis and use, we utilized a thresholding technique. We employed a threshold value of 0.5. For every individual pixel in the anticipated mask, if the pixel value is equal to or greater than 0.5, it is assigned a value of 1 to indicate the presence of a wound. Conversely, if the pixel value is below 0.5, it is assigned a value of 0, which signifies that there is no wound present. The purpose of this thresholding phase is to ensure that the output masks are binary, which is crucial for precise wound delineation and subsequent analysis.

\subsection{Evaluation Metrics}
To evaluate the efficacy of our approach, we used a variety of criteria for both segmentation and classification. These metrics give an extensive understanding of how well our model identifies wound classifications and appropriately segments wound areas.

\textbf{Accuracy:} Accuracy determines the ratio of the number of test instances that have been correctly classified out of the total number of instances Equation (\ref{eq:accuracy}). It gives a quick and simple measure of the predictive accuracy of the model in a single figure of merit.

\begin{equation}
\label{eq:accuracy}
\text{Accuracy} = \frac{\text{Number of correctly classified samples}}{\text{Total number of samples}}
\end{equation}

\textbf{F1 Score:} The F1 Score is the average of Precision and Recall values indicating that the measure takes into account both false positives and false negatives Equation (\ref{eq:f1_score}). It is especially useful when working with unbalanced datasets, in which a few classes are present.

\begin{equation}
\label{eq:f1_score}
\text{F1 score} = 2 \cdot \frac{\text{Precision} \cdot \text{Recall}}{\text{Precision} + \text{Recall}}
\end{equation}

\textbf{Precision:} Also known as "positive predictive value," refers to the proportion of pertinent cases among those recovered. It centers on the validity of positive forecasts Equation (\ref{eq:precision}).

\begin{equation}
\label{eq:precision}
\text{Precision} = \frac{\text{True Positives}}{\text{True Positives} + \text{False Positives}}
\end{equation}

\textbf{Recall:} Recall or 'sensitivity' is the proportion of pertinent cases that were retrieved. It focuses on capturing all relevant situations Equation (\ref{eq:recall}).

\begin{equation}
\label{eq:recall}
\text{Recall} = \frac{\text{True Positives}}{\text{True Positives} + \text{False Negatives}}
\end{equation}

\textbf{Dice Score:} This is a popular statistic for assessing segmentation model performance analysis, and measures the overlap between the ground truth masks and the predicted mask, indicating exact segmentation.

\section{Results}\label{sec4}

Our model's performance assessment spanned multiple metrics for classification and segmentation tasks, revealing its potential to improve accurate wound classification and segmentation.

\begin{table}[ht]
\centering
\caption{Performance of the Classification part}
\label{tab:classification_result}
\begin{tabular}{|l|c|}
\hline
Evaluation Metrics & Results \\
\hline
Accuracy  & 0.915          \\
F1 Score  & 0.907          \\
\hline
\end{tabular}
\end{table}

\begin{table}[ht]
\centering
\caption{Performance of the Segmentation part}
\label{tab:segmentation_result}
\begin{tabular}{|l|c|}
\hline
Evaluation Metrics & Results \\
\hline
Dice Score  & 0.931          \\
Precision  & 0.947          \\
Recall  & 0.865          \\
F1-Score  & 0.929          \\
\hline
\end{tabular}
\end{table}

In relation to the classification task Table \ref{tab:classification_result}, the model attained an accuracy of 0.915, successfully categorizing the four distinct wound types shown in the input images. The F1 Score for the classification task was calculated to be 0.907, reflecting solid generalization over both high and low class distributions by taking into account precision in addition to recall.

The model had a precision of 0.947 in the segmentation task Table \ref{tab:segmentation_result}, which meant that almost all pixels were correctly labeled as wound or non-wound by the predicted masks. The Dice Score of 0.931 was used to measure the extent to which the predicted and ground truth wound locations matched each other. Through a segmentation F1 Score of 0.929, it can be seen that this model is good at delineating lesions accurately too.

\section{Discussion}\label{sec5}
In the current landscape of automated wound segmentation and classification, various deep-learning models have demonstrated significant advancements. Our work aimed to expand on these foundations by suggesting a modified U-Net design with classification capacity. This section discusses how our model compares to existing methods in terms of segmentation and classification performance Table \ref{tab:comp}.

\begin{table}[ht]
\centering
\begin{tabular}{|l|l|l|l|}
\hline
\textbf{Paper} & \textbf{Model} & \textbf{Segmentation} & \textbf{Classification} \\
\hline
\cite{wang2020fully} & VGG16 & Dice: 0.810 & - \\
& & Precision: 0.839 & \\
& & Recall: 0.783 & \\
\hline
\cite{wang2020fully} & SegNet & Dice: 0.851 & - \\
& & Precision: 0.836 & \\
& & Recall: 0.865 & \\
\hline
\cite{wang2020fully} & U-Net & Dice: 0.901 & - \\
& & Precision: 0.890 & \\
& & Recall: 0.913 & \\
\hline
\cite{wang2020fully} & Mask-RCNN & Dice: 0.902 & - \\
& & Precision: 0.943 & \\
& & Recall: 0.864 & \\
\hline
\cite{wang2020fully} & MobileNetV2 & Dice: 0.903 & - \\
& & Precision: 0.908 & \\
& & Recall: 0.897 & \\
\hline
\cite{wang2020fully} & MobileNetV2 + CCL & Dice: 0.905 & - \\
& & Precision: 0.910 & \\
& & Recall: 0.899 & \\
\hline
\cite{rostami2021multiclass} & AlexNet & - & Accuracy: 0.905 \\
& & & F1-Score: 0.892 \\
\hline
\cite{chauhan2020bpbsam} & ResNet50 & - & Accuracy: 0.848 \\
& & & F1-Score: 0.778 \\
\hline
\cite{zhao2019fine} & Bi-CNN & - & Accuracy: 0.846 \\
& & & F1-Score: 0.848 \\
\hline
Our Model & Modified U-Net with Classification & Dice: 0.931 & Accuracy: 0.915 \\
& & Precision: 0.947 & F1-Score: 0.907 \\
& & Recall: 0.865 & \\
& & F1-Score: 0.929 & \\
\hline
\end{tabular}
\caption{Comparison with other models.}
\label{tab:comp}
\end{table}

\subsection{Segmentation Performance}
Rising above several state-of-the-art models, our model obtained a Dice score of 0.931, Precision of 0.947, and Recall of 0.865.  In the study of Wang et al. (2020), for example, their MobileNetV2 model, upgraded with Connected Component Labeling (CCL), Dice score of 0.905, Precision of 0.910, and Recall of 0.899 \cite{wang2020fully}. Similarly, U-Net and Mask-RCNN models reported Dice scores of 0.901 and 0.902, respectively \cite{wang2020fully}. These findings show better dependability in determining and defining wound areas as our modified U-Net not only offers better segmentation accuracy but also preserves strong precision and recall measures.

\subsection{Classification Performance}
Our model performed with an F1-score of 0.907 and an accuracy of 0.915 for classification. This beats the AlexNet-based classifier put out by Rostami et al. (2021), with an F1-score of 0.892 and an accuracy of 0.905 \cite{rostami2021multiclass}. Furthermore showing accuracy and F1-Score metrics of 0.848 and 0.778, respectively, Chauhan and Goyal's body part-specific burn severity assessment model (BPBSAM) from 2020 \cite{chauhan2020bpbsam} The outstanding performance of our model in segmentation and classification emphasizes its possibility for more exact and complete wound care treatment.

\vspace{12pt}
As we accomplished with our modified U-Net, combining segmentation and classification in one model offers several advantages. While conventional methods often concentrate on either segmentation or classification, our method addresses both. This is therefore rather crucial for wound treatment, as our model can distinguish the sorts of wounds and find their edges. Correct and timely action can help patients far better.

\section{Conclusion}
In this study, we proposed a new architecture based on U-Net that has been developed for the purpose of effective wound segmentation and classification. As for the performance indicators, our model reached the classification accuracy of 0.915 as well as the F1-score of 0.907. On segmentation, the model achieved a precision of 0.947, a Dice score of 0.931, and an F1 score of 0.929. These findings are further indicative of the accuracy of the proposed model in correctly classifying the types of wounds, as well as the level of detail it provides in terms of wound margins.

The outstanding accuracy and completeness of the model demonstrate its ability to make a huge difference in automated wound care management. Our research covers classification and segmentation with deep learning models that promise to deliver finer-grained, more efficient ways of assessing and planning treatment for wounds. This would help reduce the burden on healthcare professionals, thus leading to faster and more accurate wound care interventions.

In future work, we intend to expand our training datasets, such as by broadening them to enhance generalization in the model. The goal is also to improve the model's reliability by implementing advanced data augmentation techniques. We will have developed a mobile app that integrates our model and makes it useful in clinics or homes rather than just research settings. This initiative seeks to extend effective wound care management beyond reach, even to affordable coverage in neglected areas.

\section*{Declarations}

\subsection* {Author Contributions} Data curation, S.A. and M.A.A.N.; formal analysis, M.Z.B.J.; investigation, M.Z.B.J.; methodology, M.Z.B.J.; supervision, M.A.A.N., K.D.G., and R.G.; visualization, M.Z.B.J. and S.A.; writing---original draft, M.Z.B.J.; writing---review and editing, S.A. All authors have read and agreed to the published version of the manuscript.

%\begin{itemize}
\subsection* {Funding} A portion of the funding for this research comes from DOEd Grant P116Z220008 (1) and NSF Grant No. 2306109. Any opinions, findings, and conclusions expressed here are those of the author(s) and do not reflect the views of the sponsor(s).

\subsection* {Data Availability} The Medetec Dataset can be found at \url{https://www.medetec.co.uk/files/medetec-image-databases.html}, {accessed on 09 Feb 2024.}

\subsection*{Competing Interests} The authors declare that they have no competing interests.

\subsection*{Ethics Approval} This study did not involve human participants or animal subjects directly. The wound images used for model training and evaluation were sourced from publicly available sources.

\bibliography{sn-bibliography}
\end{document}